\documentclass[conference]{IEEEtran}
\IEEEoverridecommandlockouts
\usepackage{cite}
\usepackage{hyperref}
\usepackage{multicol}
\hypersetup{
    colorlinks=true,
    linkcolor=blue,
    filecolor=black,      
    urlcolor=blue,
}
\usepackage{xpatch}
\makeatletter
\patchcmd\@makecaption{\\}{.~}{}{\fail}
\makeatletter
\newcommand{\linebreakand}{%
  \end{@IEEEauthorhalign}
  \hfill\mbox{}\par
  \mbox{}\hfill\begin{@IEEEauthorhalign}
}
\usepackage{tabularx,booktabs}
\usepackage{amsmath,amssymb,amsfonts}
\usepackage{algorithmic}
\usepackage{graphicx}
\usepackage{textcomp}
\usepackage{xcolor}
\def\BibTeX{{\rm B\kern-.05em{\sc i\kern-.025em b}\kern-.08em
    T\kern-.1667em\lower.7ex\hbox{E}\kern-.125emX}}
\begin{document}

\title{Nutritional Profile Estimation in Cooking Recipes }

\author{
\IEEEauthorblockN{1\textsuperscript{st} Jushaan Kalra}
\IEEEauthorblockA{\textit{Dept. of Computer Engineering} \\ 
\textit{Delhi Technological University}\\
New Delhi, India \\ jushaan18@gmail.com
}
\and
\IEEEauthorblockN{1\textsuperscript{st} Devansh Batra}
\IEEEauthorblockA{\textit{Dept. of Information Technology} \\ 
\textit{NSUT}\\
New Delhi, India \\
devanshb.it.17@nsit.net.in}
\linebreakand

\IEEEauthorblockN{2\textsuperscript{nd} Nirav Diwan}
\IEEEauthorblockA{\textit{Dept. of Computer Science and Engineering} \\
\textit{IIIT-Delhi}\\
New Delhi, India \\
nirav17072@iiitd.ac.in}
\and

\IEEEauthorblockN{Ganesh Bagler}
\IEEEauthorblockA{\textit{Center for Computational Biology} \\
\textit{IIIT-Delhi}\\
New Delhi, India \\
bagler@iiitd.ac.in}
}
\maketitle

\begin{abstract}
The availability of an accurate nutrition profile of recipes is an important feature for food databases with several applications including nutritional assistance, recommendation systems, and dietary analytics. Often in online databases, recipes are obtained from diverse sources in an attempt to maximize the number of recipes and variety of the dataset. This leads to an incomplete and often unreliable set of nutritional details. We propose a scalable method for nutritional profile estimation of recipes from their ingredients section using a standard reliable database for the nutritional values. Previous studies have testified the efficiency of string-matching methods on small datasets. To demonstrate the effectiveness of our procedure, we apply the proposed method on a large dataset, RecipeDB~\cite{b1}, which contains recipes from multiple data sources, using the United States Department of Agriculture Standard Reference (USDA-SR) Database as a reference for computing nutritional profiles.  We evaluate our method by calculating the average error across our database of recipes (36 calories per serving) which is well within the range of errors attributable to physical variations.
\end{abstract}


\section{Introduction}
Estimating the nutritional profile of a cooking recipe is a challenging problem. While there is no dearth of web-based services that provide recipes, their cooking instructions along with ingredient details, pertaining to a wide range of cuisines across the world, their nutritional profiles are not easily available. Here, we propose a Named Entity Recognition(NER)-based strategy for extracting different elements of recipes and to compute the nutritional profile of a recipe by mapping them to their USDA nutritional description.

Several methods for the calculation of nutritional values of a cooked meal have been proposed. The most accurate method~\cite{b1} for this calculation employs chemical analysis. Since this method is applied on the cooked meal, it does not lead to any untoward errors.
However, this analysis is not feasible for large datasets of recipes from online resources, since user-uploaded recipes tend to be extremely noisy and without a standard format for storing data. Furthermore, it is not practical to conduct chemical analysis on every recipe, since they may number in hundreds of thousands. Through the course of our research, we collected more than 100,000 recipes from one source alone and hence we sought for more scalable methods.

An alternative approach is mentioned in~\cite{b2} where food images are used to calculate calorie contents.
Such methods do not provide accurate results suitable for academic research. Since these methods also look for the presence of particular ingredients within food images which are themselves available more accurately in the recipe text, we focus on methods that use the text content itself.

The approach we adopted is aligned with the one mentioned in~\cite{b3} which assumes that the sum total of nutrition of ingredients in a particular recipe can be approximated for the nutritional profile of the recipe. This simplifies our problem statement since we can now calculate the nutritional value of ingredients from nutritional composition tables, and their sum would give us our required nutritional values.

It has been observed~\cite{b4} that more accurate results would be obtained if nutritional yield due to cooking is taken into account, but, there is no such consolidated resource for yield values as they differ with ingredient, cooking time and other variable features. Without the knowledge of these variables, it is difficult to estimate the nutritional profile of the recipe with the above method.

\begin{figure*}
    \begin{center}
   \includegraphics[height=12cm]{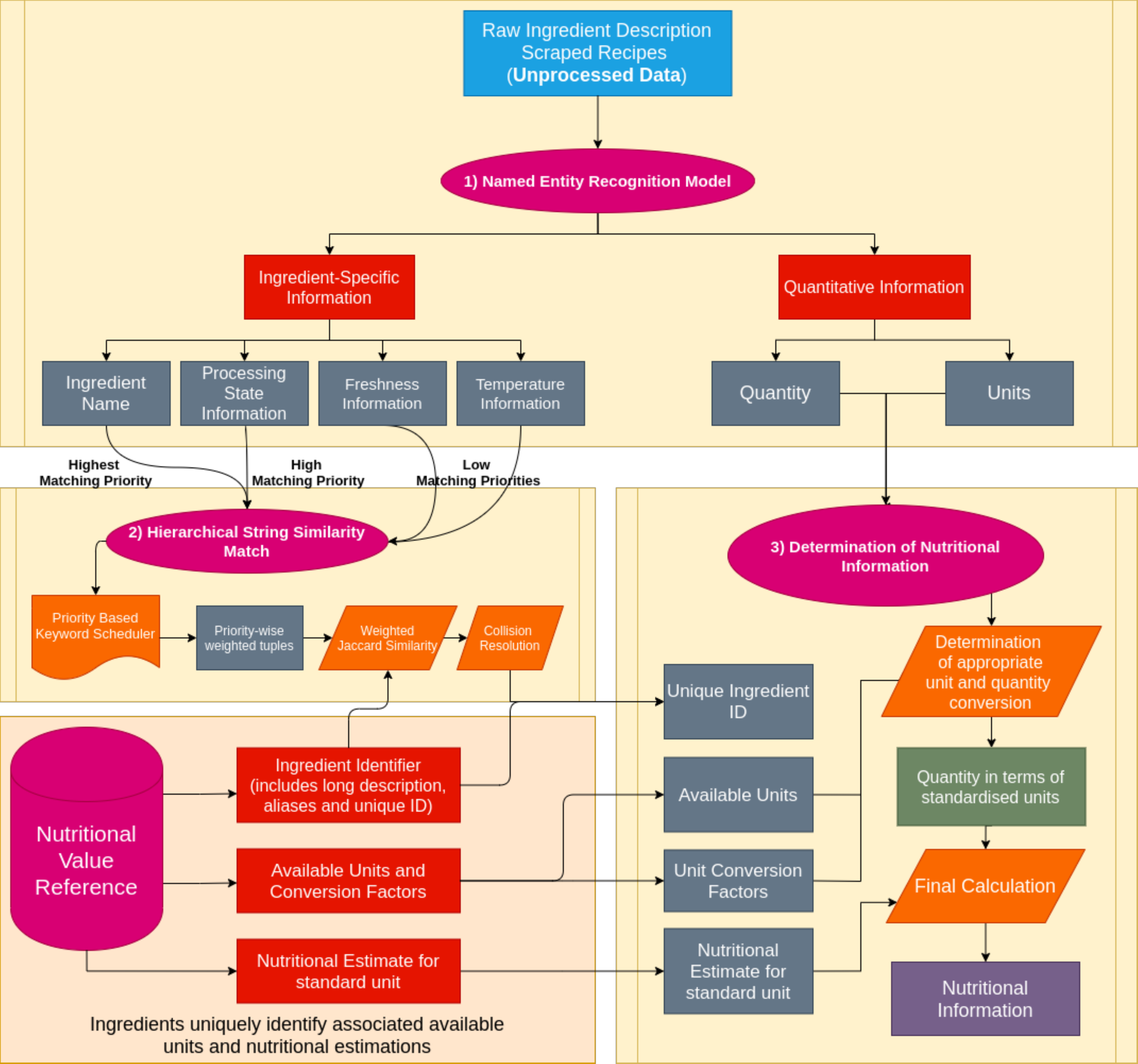}
     \end{center}
  \caption{System Architecture}
  \label{System_Architecture}
\end{figure*}

Hence the task is to form a complete nutritional composition table for all ingredients in our recipes since the nutritional value of a recipe is the sum total of the nutritional value of its constituent ingredients. To calculate this, we propose a three-step approach-- Ingredient Data Mining (Section~IIA), Ingredient Name Matching (Section~IIB) and Unit Matching (Section~IIC) which together give us the required nutritional profile. Refer to Figure~\ref{System_Architecture} for an overview of the system architecture used.

\section{Calculating Nutritional Value of Recipes}
\subsection{Ingredient Data Mining}


\def\arraystretch{1.5}%
\begin{table*}
 \caption{Ingredient Tags Extraction}
\label{table:ner}
\begin{tabularx}{\textwidth}{@{}l*{10}{c}c@{}}
\toprule
Ingredient Phrase & Name & State & Quantity & Unit & Temperature & Dry/Fresh & Size  \\ 
\midrule
1/2 lb lean ground beef   & beef & ground lean & 1/2 & lb & &  &   &  \\ 
1 small onion , finely chopped & onion & chopped  & 1 & &  &  & small \\
1 hard-cooked egg , finely chopped  & egg & hard-cooked chopped  & 1 & &  &  & \\
1 tablespoon fresh dill weed & dill weed & & 1 & tablespoon &  & fresh & \\
1/2 teaspoon salt   ,freshly ground & salt & & 1/2 & teaspoon &  &  & \\
1/8 teaspoon black pepper,minced & black pepper & & 1/8 & teaspoon &  & & \\
3/4 cup butter or 3/4 cup margarine , softened & butter & softened & 3/4 & cup &  &  & \\
2 cups all-purpose flour & purpose flour &  & 2 & cups &  &  & \\
1 teaspoon salt & salt &  & 1 & teaspoon &  &  & \\
1/2 cup low-fat sour cream & cream & sour low fat  & 1/2 & cup &  &  & \\
1 egg yolk  & egg yolk & & 1 & &  &  & \\
1 tablespoon cold water & cold water & & 1 & tablespoon & cold  &  & \\
\bottomrule
\end{tabularx}
\end{table*}

\def\arraystretch{1.7}
\begin{table}[htbp]
\caption{Examples of Food Description in USDA-SR Database}
\begin{center}
\begin{tabular}{|c|c|}
\hline
\textbf{S.No}& \textbf{Description} \\
\cline{1-2} 

\hline
1 & Butter, salted \\
\hline
2 & Butter, whipped, with salt \\
\hline
3 & Butter, without salt \\
\hline
4 & Cheese, blue \\ 
\hline
5 & Cheese, cottage, creamed, large or small curd \\
\hline
6 & Cheese, mozzarella, whole milk \\ 
\hline
7 & Milk, reduced fat, fluid, 2\% milkfat,\\ & with added vitamin A and vitamin D \\ 
\hline
8 & Milk, reduced fat, fluid, 2\% milkfat, with added \\ 
& nonfat milk solids and vitamin A and vitamin D \\
\hline
9 & Milk, reduced fat, fluid, 2\% milkfat, \\ & protein fortified, with added vitamin A and vitamin D \\
\hline
10 & Milk, indian buffalo, fluid \\
\hline
11 & Milk shakes, thick chocolate \\
\hline
12 & Milk shakes, thick vanilla \\
\hline
13 & Yogurt, plain, whole milk, 8 grams protein per 8 ounce \\
\hline
14 & Yogurt, vanilla, low fat, 11 grams protein per 8 ounce \\
\hline
15 & Egg, whole, raw, fresh \\
\hline
16 & Egg, white, raw, fresh \\ 
\hline
17 & Egg, yolk, raw, fresh \\
\hline
18 & Apples, raw, with skin \\
\hline
19 & Apples, raw, without skin \\
\hline


\end{tabular}
\label{table:long-descriptions}
\end{center}
\end{table}

\def\arraystretch{2}%
\begin{table*}
\caption{Comparison of Inferences from Modified Jaccard Index with Vanilla Jacard Index}
\label{table:jaccard}
\begin{tabularx}{\textwidth}{@{}l*{10}{c}c@{}}
\toprule
\textbf{Ingredient Phrase} & \textbf{Ingredient Name} & \textbf{Food Desc. Inferred W/ Modified JI} & \textbf{Food Desc. Inferred W/ Vanilla JI}  \\
\midrule
1 cup red lentil &  red lentils & Lentils, pink or red, raw & Cherries, sour, red, raw \\
\hline
1 roma tomato , quartered &  roma tomato & Soup, tomato beef with noodle, & Soup, tomato, canned \\
 &   & canned, condensed & condensed\\
\hline
1/4 teaspoon ground coriander &  coriander & Coriander (cilantro) leaves, raw & Spices, coriander leaf, dried \\
\hline
2 tablespoons tomato paste &  tomato paste & Tomato products, canned, paste, without salt added & Soup, tomato, canned, condensed \\
\hline
1 1/4 cups vegetable broth &  vegetable broth & Soup, vegetable with beef broth, canned, condensed & Soup, vegetable broth, ready to serve \\
\hline
1 can fava beans &  fava beans & Broadbeans (fava beans), mature seeds, raw & Beans, fava, in pod, raw \\
\hline
1 teaspoon ground cayenne pepper &  cayenne pepper & Spices, pepper, red or cayenne & Spices, pepper, black \\
\hline
1 whole chicken with giblets & chicken with giblets & Chicken, broilers or fryers, meat and skin & Fast foods, quesadilla, with chicken \\
patted dry and quartered & & and giblets and neck, raw & \\
\hline
2 tablespoons sesame seeds &  sesame seeds & Salad dressing, sesame seed dressing, regular & Seeds, sesame seeds, whole, dried \\
\hline
1 teaspoon ground coriander &  coriander & Coriander (cilantro) leaves, raw & Spices, coriander leaf, dried \\
\bottomrule
\end{tabularx}
\end{table*}


We utilize the data available from RecipeDB~\footnote{https://cosylab.iiitd.edu.in/recipedb}~\cite{b1} which contains 118,071 recipes from  
AllRecipes~\footnote{https://www.allrecipes.com} and {FOOD.com}~\footnote{https://www.food.com}.
In order to estimate the nutritional profile of a recipe, we need to obtain all the ingredients used in a recipe and their corresponding quantities, units and/or size and other useful information such as processing state (ground, thawed, etc.), temperature and dryness.

Consider the recipe - Piroszhki (Little Russian Pastries)~\footnote{https://www.food.com/recipe/piroszhki-little-russian-pastries-486254}. The Table~\ref{table:ner} shows the outputs of our Named Entity Recognition approach on twelve ingredient phrases. We note that for example, in the table, ``1 small onion, finely chopped'' contains the entire information that we require to calculate the ingredient's nutritional value, we only need the data in a structured format in order to estimate the nutritional value of the recipe. 

We propose a Named Entity Recognition System to train the model to infer the following tags-- NAME, STATE, UNIT, QUANTITY, TEMP, DRY/FRESH, SIZE. We manually tagged a corpus of 6612 ingredient phrases and tested the model on a test set of size 2188 ingredient phrases. In order to include ingredient phrases of large diversity in our training and testing set, we utilized Parts of Speech Tagging to form vectors representing each ingredient phrases. A vector representing an ingredient phrase would be defined by the frequency of the tag in the ingredient phrase. 
We then proceeded to cluster the obtained vectors. The ingredient phrases were chosen for the training and testing set by selecting a subset of ingredient phrases from each cluster. We trained our model using Stanford Named Entity Recognition Model~\cite{b5}.
The model obtained an F1 score of 0.95 on the test set validated by 5-fold cross validation.

The next two sections explain how ingredient descriptions from the recipes are mapped to ingredient description from the nutritional database (Section IIB) and secondly, how the units of the ingredients are mapped to one of the available units present in the nutritional database (Section IIC).

\subsection{Closest Description Annotation Using String Similarity Matching}\label{AA}
In order to accurately map ingredient names to food descriptions in the Standard Reference database, we carefully looked for patterns in food description strings that might help us select the best possible description. 

\textbf{(a)} It can be observed that the descriptions in the USDA-SR database are comma-separated terms with a decreasing degree of importance associated with each consequent term. Consider all descriptions from the food description column of Table~\ref{table:long-descriptions}. The first term is significant for matching. Hence, Butter, Cheese, Milk, Milk shakes, Yogurt, Egg and Apples occupy the highest priority for finding a match within the ingredient description.


\textbf{(b)} The high priority terms include both singular and plural forms of nouns. They must be lemmatized before matching. For this purpose, we used the NLTK library's WordNet Lemmatizer~\cite{b6}. Stemmers, although computationally less expensive, were not found to be useful for this purpose because of their high aggression.

\textbf{(c)} The Ingredient Name ``Egg whites'' best matches with the description ``Egg, white, raw, fresh'' whereas ``Whole eggs'' best match ``Egg, whole, raw, fresh''. The sequence of terms may be different in both the strings being considered. To tackle this, we use a modified form of Jaccard Index~\cite{b7} as the metric for the similarity between the two strings. The modified Jaccard distance has been explained in \textbf{(e)}.

\textbf{(d)} Another observation is that the comma-separated terms in later portions of the food description are more likely to match with the State, Temperature and Freshness of the ingredient. Therefore, we match the whole description along with the State, Temperature and Freshness entities derived from our NER pipeline.

\textbf{(e)} We would like to prioritize the mapping of maximum terms from the Ingredient phrase rather than the food description using a vanilla Jaccard Index. This is because a lot of food descriptions include additional details unspecified in the ingredient description.
Assume A and B are the set of words formed after preprocessing the Ingredient Phrase and Food Description respectively by lemmatization, stop-word removal and uniform casing, and $\vert$A$\vert$ and $\vert$B$\vert$ are the number of words in these sets. Similarly $\mid A \cup B \mid$ gives the number of words in the union of the sets of words in strings A and B and $\mid A \cap B \mid$ gives the number of words in the intersection of the sets of words in strings A and B.

\[
    J(A,B) = \frac{\mid A \cap B \mid}{\mid A \cup B \mid}
\]

For a lot of descriptions, $\vert$B$\vert$ is extremely large, consider food description for serial numbers 7, 8, 9, 13, 14. The denominator in Jaccard distance increases with an increase in $\vert$B$\vert$. This leads to a bias against large strings. However, it is only essential to match the maximum number of terms from the Ingredient Phrase. So, we use $\vert$A$\vert$ as the denominator for our modified Jaccard Matching Index. 
\[
    J^{\ast}(A,B) = \frac{\mid A \cap B \mid}{\mid A \mid}
\]

This modification removes bias against large strings of detailed food descriptions without which ``skimmed milk'' would match with ``Milk Shakes, thick chocolate'' completely disregarding more accurate ``Milk, reduced fat, fluid, 2\% milkfat, protein fortified, with added vitamin A and vitamin D''. This bias was found to be highly significant with 227 out of 1000 randomly sampled ingredient phrases from RecipeDB having a different match. Some observed inferences from the two metrics have been shown in Table ~\ref{table:jaccard}. It can clearly be seen that the vanilla Jaccard Index is biased towards shorter, less detailed food descriptions, proving that the Modified Jaccard Index is more suitable for the task.  

\textbf{(f)}  We also need to account for the presence of negation terms. ``unsalted butter'' should match with ``Butter, without salt''. To achieve this, we replaced all negation terms and prefixes (like ``un'' in unsalted) to ``not''. Following this, the preprocessing mentioned above is applied and the Ingredient Phrase and Description respectively become ``not salt butter'' and ``butter not salt'' leading to a perfect Jaccard Match.

\textbf{(g)} Many food descriptions include ``raw'' which is analogous to uncooked state or cases where State is not mentioned. So we include a provision to match an additional word when ``raw'' occurs in the description and no State is identified. This way, ``apple'' matches with ``Apples, raw, with skin''.

\textbf{(h)}  We also store the sequence numbers (priority) of terms along with the Modified Jaccard Index score. This helps us in resolving collisions such as matching “apple” with “Apples, raw, with skin” and “Babyfood, apples, dices, toddler”. The first description is chosen because it contains the matching word “Apples” in the first term (having a higher priority than occurring at later indices). This follows the line of reasoning as explained in
\begin{figure*}
  \includegraphics[width=2\columnwidth]{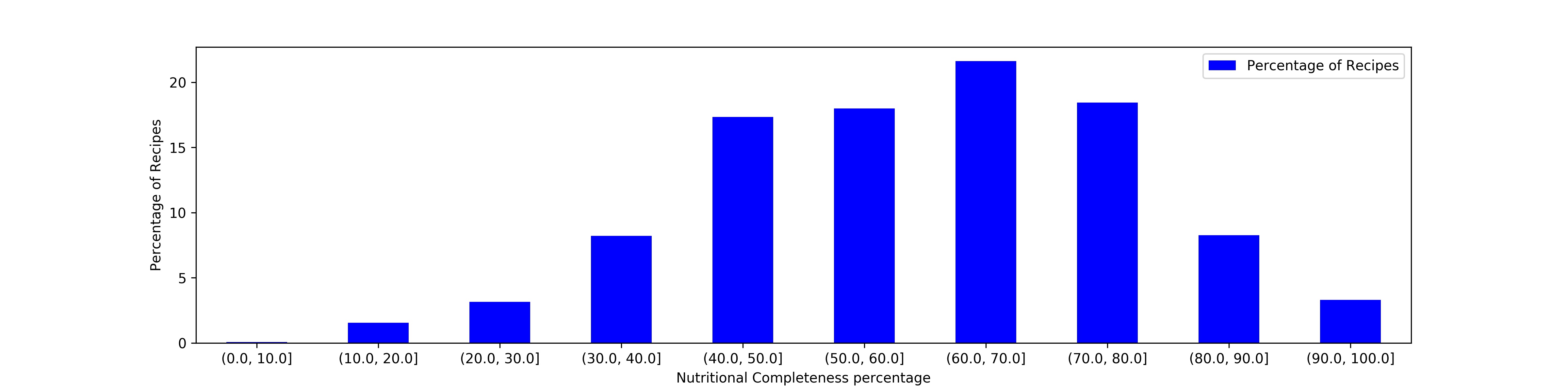}
  \caption{Percentage mapping of recipes to their nutritional profile}
  \label{nutri_map}
\end{figure*}
\textbf{(a)}

\textbf{(i)} In case collisions cannot be resolved through Jaccard Index and Sequential Priority, we simply take the first match. This is useful because of the way the descriptions have been indexed within USDA-SR Database. ``Apple'' matches with the more desirable ``Apples, raw, with skin'' as opposed to ``Apples, raw, without skin'' and ``Eggs'' match with ``Egg, whole, raw, fresh'' rather than the other variants which are more suitable for ``egg whites'' and ``egg yolk'' even though they both return the same Jaccard Index and priority score with ``Egg''.

\vspace{10pt}

\subsection{Units Matching and subsequent Nutrition calculation}\label{AB}

Once an ingredient from a recipe has been matched to its corresponding food description in the nutrition database, we match the unit corresponding to this ingredient from our nutritional database.
Unfortunately, in this case string matching techniques would not be satisfactory since we have a fixed number of possible units with a fixed format and applying heuristics similar to Section II B might give unwanted results due to incorrect matching of strings. Furthermore, the units provided in the nutritional database may not be enough. For example - our dish requires ``1 teaspoon of butter'' whereas, in the USDA database, we do not have teaspoon in available units for butter. On top of that some of the units are not clean, e.g. `pat (1" sq, 1/3" high)' was one of the units in the USDA-SR Database. See Table~\ref{fig:tab2}

Similar problems exist in the units used for the dish. Adding to that, we may have different aliases referencing the same unit in our data, e.g. `tablespoon' and `tbsp' refer to the same unit and so do `pound' and `lb'.

To circumvent these problems, we applied WordNet Lemmatization using NLTK library~\cite{b6} on all the units present in our recipes and USDA-SR database then took the first word and applied Regular Expression(regex) to obtain a cleaner version containing only alphabets (this helps us to ignore noise and keep relevant part like taking pat out of `pat (1" sq, 1/3" high)').
Furthermore, standard units were defined for units where aliases were present, for example, tbsp and tablespoon both now represent the standard unit tablespoon.
To deal with the case where a unit could not be found, measurement conversion tables were created with detailed conversions between units on the basis of volume using measurements mentioned in~\cite{b8}. These tables were used to check for the missing units. The tables mention conversions such as `1 cup' is equivalent to `16 tbsp' and `48 tbsp' and so on.

Table~\ref{fig:tab2} shows the data after lemmatizing and data cleaning. For butter, the units `cup' and `tablespoon' are present, but `teaspoon' is not. Hence, we can add teaspoon as a unit since the ratio of volume of a cup and a teaspoon is constant. The sizes small, medium and large were mentioned in both, the nutritional database and recipes as units. All 3 were considered equivalent because of ambiguity between sizes.
\begin{table}[htbp]
\caption{ingredient and unit relations}
\begin{center}
\begin{tabular}{|c|c|c|c|c|c|}
\hline
\textbf{ingredient}& \textbf{seq} & \textbf{amount}  & \textbf{unit} & \textbf{grams} &\textbf{gram\_per\_amount} \\
\cline{2-4}

\hline
Butter,salted &  1 & 1.0 & pat & 5.0 & 5.0\\
\hline
Butter,salted & 2 & 1.0 & tbsp & 14.2 & 14.2\\
\hline
Butter,salted & 3 & 1.0 & cup & 227.0 & 227.0\\
\hline
Butter,salted & 4 & 1.0 & stick & 113.0 & 113.0\\
\hline

\end{tabular}
\label{fig:tab2}
\end{center}
\end{table}








After cleaning and lemmatizing the ingredient unit and quantity, we calculate the nutrition profile of each ingredient by merging the recipe data and nutrition data on the unit and multiplying the nutrition profile by the quantity of the ingredient. Quantities were preprocessed to match a specific numerical value. `2-4' was averaged to 3, `2 1/2' was converted to 2.5 and so on. 

In certain cases NER did not detect units, in that scenario we searched the ingredient phrase for known units and if found they were updated. Certain ingredient phrases included statements such as `500 g or 1 cup' which the NER wrongly detected as `500 cups'. This was dealt in a semi-automated manner by putting a threshold on the quantity per unit. Finally, wherever a unit was still not present, the most frequent unit for that particular ingredient was used. This works well to maintain consistency in the data since we have a lot of units corresponding to each ingredient, but only a few of them are dominant, e.g. for garlic, if the unit was not detected, it would most probably be clove which we obtained using the above-mentioned heuristic.


\section{Results}
Using heuristics mentioned in II-B we were able to match 94.49\% of the unique ingredients from the recipes, with the rest remaining unmapped from the USDA dataset. To assess the validity of the jaccard matching, the 5000 most frequent ingredients+states were manually matched with the USDA dataset, out of which 3580 were deemed to be correct matches, the rest had a better match available in the dataset (accuracy of 71.6\%). It is important to note here that USDA has a lot of similar ingredients with little variation as is evident from Table~\ref{table:long-descriptions}, so while jaccard similarity does not always give the best match, it almost always gives one of the suitable matches from our database.

To further probe how many ingredients along with their units could be mapped to the USDA dataset, we analyzed percentage mapping of recipes to their nutritional profile in terms of the percentage of ingredients in a recipe getting mapped to their USDA nutritional profiles (Figure~\ref{nutri_map}). It indicates that the protocol implemented could successfully map a significant proportion of ingredients to their nutritional profiles thereby contributing to the accuracy of the estimated nutritional profiles of recipes. The figure also indicates that the main problem lies in matching the units of ingredients to appropriate units in the USDA dataset, especially when some units are not mentioned in the nutritional database itself.

Calorie information from All Recipes was extracted and used as a baseline to evaluate our results. The nutritional profiling of recipes at AllRecipes was done by outsourcing it to a reliable third-party. We consider this as the gold standard for the evaluation of our estimated nutritional profiles. We selected data for which we had 100\% mapping of ingredients with their nutritional values, and had clean, well-defined servings. This resulted in 2482 recipes. This was done because while our recipe dataset has a global coverage, spanning 26 regional cuisines, the sample food composition table that was used mostly contained details of ingredients used in the United States. For e.g. `garam masala'- a spice used in Indian dishes is not an ingredient present in the dataset. Because of these region-centric ingredients, some ingredients were not mapped. Incorporation of other data as mentioned in Food and Agricultural Organisation of the United Nation\footnote[5]{http://www.fao.org/infoods/infoods/tables-and-databases/en/} would help in improving the results.

For the chosen 2482 recipes we observed an average per serving error of 36.42 calories. To put this in the context in terms of the ingredient used in Table~\ref{fig:tab2}, (butter, salted), 1 teaspoon of it is equivalent to 35 calories. This is well within our scope of error since some calorie content would differ based on the user, cooking time and utensils used in cooking the dish.

\section{Conclusions}
 We use NER with Jaccard Similarity and Unit Mapping on a large database containing more than 118,000 recipes to provide accurate estimates of nutritional profiles despite extremely noisy and varied data. We show that the proposed protocol is robust, compatible with any nutritional database, easily replicable and solves one of the foremost problems with dietary analysis and food recommendation systems. We provide the code on Github\footnote[6]{https://github.com/cosylabiiit/Nutritional-Estimation-In-Recipes}. We would like to highlight that our system provides a good `estimate' for the nutritional value of food and as nutritional composition tables get updated, our heuristics will give better results without any changes. 
 
\section*{Acknowledgment}
J.K. and D.B. contributed equally towards the the work done for this research article. J.K., D.B. and N.D. thank IIIT-Delhi and Complex Systems Laboratory for the Summer Research Internship. G.B. thanks IIIT-Delhi for the computational facilities. The authors thank Rudraksh Tuwani for suggesting the use of NER as part of computational protocol.

\vspace{12pt}
\color{red}
\end{document}